# Convective flux analysis on the propagation mechanism of oblique detonation waves

Yunfeng Liu[1,2]

1. Institute of Mechanics, Chinese Academy of Sciences, Beijing 100190, China

2. School of Engineering Science, University of Chinese Academy of Sciences, Beijing 100049, China

liuyunfeng@imech.ac.cn

**Abstract：** The aim of this study is to investigate the propagation mechanism of oblique detonation waves using the vector flux analysis method through numerical simulations. A two-dimensional numerical study is conducted on stoichiometric hydrogen-air oblique detonation waves based on the conservative Euler equations and a one-step global chemical reaction model. The wedge angle is 25°, with a freestream static temperature of 851.5 K, velocity of 2473.4 m/s, and pressure of 42.5 kPa. The motion mechanism of transverse waves is analyzed using the vector flux method. The results show that the oblique detonation front consists of three regions: an induction zone, an overdriven detonation zone, and a transverse-wave region. Under different activation energies, only either upward-propagating or downward-propagating transverse waves exist on the oblique detonation front; the two do not occur simultaneously. At low activation energy, downward-propagating transverse waves dominate, whereas at high activation energy, upward-propagating transverse waves appear.

**Keywords：** oblique detonation wave; transverse wave; vector flux analysis method; ignition delay time; auto-ignition

## 1. Introduction

When a hypersonic premixed combustible gas flows over a wedge or cone surface, an oblique shock wave is generated. Behind the oblique shock, a deflagration wave is induced, and as this deflagration wave propagates downstream and interacts with the oblique shock, an oblique detonation wave (ODW) attached to the solid wall is formed [1-3]. An oblique detonation engine (ODE), which organizes combustion through oblique detonation, offers significant advantages such as self-pressurization, rapid heat release, and high thermal-cycle efficiency [4,5]. In recent years, ODEs have achieved breakthrough experimental progress [6-11], demonstrating significant potential for air-breathing hypersonic propulsion and refocusing research attention on the classical fluid-dynamic phenomenon of oblique detonation. The flow mechanism of ODWs is a fundamental scientific problem in combustion research.

Stable combustion is the prerequisite for the reliable operation of air-breathing hypersonic vehicles [12,13]. However, instability is an intrinsic characteristic of detonation waves [14-16]. For ODWs in confined configurations, instabilities stem from detonation-wall reflections and the resulting turbulent boundary-layer separation [17,18]. For ODWs in open configurations, instabilities mainly manifest in two forms: (1) large-scale instability of the macroscopic wave structure [19-21], and (2) the formation of cellular structures due to wave-front instability [22-24]. Choi et al. [19] numerically reproduced Stanford's oblique detonation experiments conducted in an expansion tube [25,26], revealing that when the wedge angle exceeds the maximum value predicted by the shock polar, the flow field depends on the ratio of flow-characteristic time to chemical-reaction characteristic time. Simulations further showed that ODWs in expansion tube environments are inherently unstable; however, due to the short test durations, oscillatory flow structures could not be observed experimentally. The instability of the macroscopic wave system is influenced by inflow conditions [20,21], wedge angle [27,28], and wedge length [19], and its mechanism is relatively well understood.

In contrast, wave-front instability and its associated small-scale cellular structures are far more complex. Choi et al. [29], by solving the Euler equations with a one-step global reaction model, investigated the effects of activation energy and grid resolution on the simulation of wave-front cellular structures. They found that at low activation energy, the wave front remains smooth even when the grid is refined; at high activation energy, the wave front becomes unstable and develops cellular patterns. However, when the grid is too coarse, the wave front still appears smooth, and only sufficiently fine grids are able to capture the cellular structures induced by wave-front instability. In addition to activation energy, Yang et al. [30] reported that the pre-exponential factor has a significant impact on wave-front stability, with larger values promoting instability. Furthermore, increasing heat release also tends to trigger detonation-front instability [31].

For unstable oblique detonation fronts, transverse shocks can be clearly observed. Verreault et al. [32] conducted unsteady numerical simulations and found that left-running and right-running transverse waves sustain the structure of the unstable front.

Researchers often use the overdrive factor as a criterion for determining front instability, with higher overdrive levels suppressing instability. Verreault et al. [32] argued that the critical overdrive factor for oblique detonation is approximately 1.8, which is nearly identical to the critical value required to maintain a stable one-dimensional piston-driven detonation, suggesting that a two-dimensional steady-state oblique detonation is fundamentally equivalent to a one-dimensional unsteady detonation. Grismer et al. [33] reached a similar conclusion, finding that the overdrive level needed to stabilize an oblique detonation front is slightly greater than the critical value predicted by one-dimensional theory. However, Teng et al. [34] showed that even at relatively high overdrive levels, the detonation front can still become unstable. This indicates that although the overdrive factor can qualitatively describe the onset of instability in oblique detonation fronts, it cannot fundamentally explain their underlaying instability mechanism. A deeper understanding and explanation of the intrinsic instability of oblique detonations is therefore essential for stabilizing oblique detonation waves under complex inflow and boundary conditions.

Detonation represents a fundamentally multi-scale phenomenon arising from the tight coupling between a shock front and a chemical reaction zone. The resulting flow structures are highly intricate, and interpretation based solely on macroscopic thermodynamic fields-such as pressure, density, temperature, and velocity-offers limited physical clarity. Vector fluxes, in contrast, constitute the most elementary and irreducible building blocks of CFD formulations. Leveraging this framework, prior work by the authors has successfully elucidated the deflagration-to-detonation transition (DDT), captured the intrinsic instability dynamics of one-dimensional unstable detonations, and quantitatively linked one-dimensional unsteady detonations to two-dimensional cellular structures [35-37]. In the present study, the vector-flux analysis methodology is extended to examine oblique detonation propagation in an unconfined environment. This approach enables direct interrogation of the underlying momentum and energy transport processes, revealing new physical insights into the stability characteristics and wave-front dynamics of oblique detonation waves.

## 2. Formulation and Numerical Methodology

The governing equations consist of the two-dimensional conservative Euler equations coupled with a one-step global reaction model. Viscous effects, molecular diffusion, and heat conduction are neglected.

$$\frac{\partial U}{\partial t} + \frac{\partial F}{\partial x} + \frac{\partial G}{\partial y} = S \tag{1}$$

$$U = \begin{pmatrix} \rho \\ \rho u \\ \rho v \\ \rho e \\ \rho Z \end{pmatrix} \quad F = \begin{pmatrix} \rho u \\ \rho u^2 + p \\ \rho uv \\ (\rho e + p)u \\ \rho uZ \end{pmatrix} \quad F = \begin{pmatrix} \rho v \\ \rho uv \\ \rho v^2 + p \\ (\rho e + p)v \\ \rho vZ \end{pmatrix} \quad S = \begin{pmatrix} 0 \\ 0 \\ 0 \\ 0 \\ \dot{\omega} \end{pmatrix} \tag{2}$$

$$e = \frac{p}{(\gamma - 1)\rho} + \frac{1}{2}\left(u^2 + v^2\right) + Zq \tag{3}$$

$$p = \rho RT \tag{4}$$

$$\dot{\omega} = -K\rho Z \exp\left(-\frac{CE_a(\gamma - 1)}{RT}\right) \tag{5}$$

$$\gamma(Z) = \frac{\gamma_1 R_1 Z / (\gamma_1 - 1) + \gamma_2 R_2 (1 - Z) / (\gamma_2 - 1)}{R_1 Z / (\gamma_1 - 1) + R_2 (1 - Z) / (\gamma_2 - 1)} \tag{6}$$

$$R(Z) = R_1 Z + R_2 (1 - Z) \tag{7}$$

Here, $\rho$, $p$, $T$, $u$, $v$, $e$, $q$, $R$, $\gamma$, $Z$ denote the density, pressure, temperature, flow velocity, specific total internal energy, heat release per unit mass, gas constant, specific heat ratio, and reaction progress variable, respectively. In the one-step global reaction model, $\dot{\omega}$ represents the mass production rate due to chemical reaction, $K$ is the pre-exponential factor, and $E_a$ is the activation energy. In the numerical simulations, the activation energy is varied by adjusting the parameter $C$ in Eq. (5). A variable specific heat ratio and a variable gas constant are adopted for the detonation products to better represent the thermodynamic properties of the reaction mixture. Subscripts 1 and 2 in Eqs. (6) and (7) denote the reactants and products, respectively.

Eqs. (8)-(15) describe the evolution of the conservative variables within each time step under an explicit time-marching scheme. The pressure increment Δp at a grid point during a single time step results from the algebraic sum of three contributions: a convective term, a kinetic-energy term, and a heat-release term. Although the heat-release term is always positive within a time step, chemical energy release does not necessarily produce a pressure rise. Whether combustion heat release results in a pressure increase depends on whether the convective contribution during that time step corresponds to a compression wave or a rarefaction wave.

$$U^{n+1} = U^n - \frac{\partial F^n}{\partial x}\Delta t - \frac{\partial G^n}{\partial y}\Delta t + S^n \Delta t \tag{8}$$

$$\Delta(\rho e) = (\rho e)^{n+1} - (\rho e)^n = -\frac{\partial(\rho e + p)u}{\partial x}\Delta t - \frac{\partial(\rho e + p)v}{\partial y}\Delta t \tag{9}$$

$$\rho e = \frac{p}{(\gamma - 1)} + \frac{1}{2}\rho\left(u^2 + v^2\right) + \rho Zq \tag{10}$$

$$\Delta p = p^{n+1} - p^n = \left(\Delta(\rho e) - \Delta\left(\frac{1}{2}\rho\left(u^2 + v^2\right)\right) - \Delta(\rho Zq)\right)(\gamma - 1) \tag{11}$$

$$\Gamma = \Delta(\rho e) = -\frac{\partial(\rho e + p)u}{\partial x}\Delta t - \frac{\partial(\rho e + p)v}{\partial y}\Delta t \tag{12}$$

$$\Delta Q = -\Delta(\rho q Z) = \rho^n Z^n q - \rho^{n+1} Z^{n+1} q \tag{13}$$

$$\Delta K = -\Delta\left(\frac{1}{2}\rho\left(u^2 + v^2\right)\right) = \frac{1}{2}\rho^n\left(\left(u^n\right)^2 + \left(v^n\right)^2\right) - \frac{1}{2}\rho^{n+1}\left(\left(u^{n+1}\right)^2 + \left(v^{n+1}\right)^2\right) \tag{14}$$

$$\Delta p = (\Gamma + \Delta K + \Delta Q)(\gamma - 1) \tag{15}$$

The freestream conditions correspond to an oblique detonation engine operating at a flight Mach number of 9 and an altitude of 30 km. The inflow is assumed to pass through two equal-strength compression shocks in the inlet. In the numerical simulation, the exit of the inlet is modeled as a stoichiometric hydrogen-air mixture, with static temperature $T$ =851.5K, velocity $u$ =2473.4m/s, pressure $p$ =42.5kPa, the wedge angle is 25°. The kinetic parameters of the one-step global reaction model for the hydrogen-air mixture follow Refs. [38,39]. A second-order ENO scheme is employed for spatial discretization together with a third-order TVD Runge-Kutta method for time advancement. The convective fluxes are computed using the Steger-Warming flux-vector-splitting approach. The entire computational domain is discretized with a uniform grid of 20μm. Freestream boundary conditions are imposed on the left and upper boundaries, a slip-wall (mirror reflection) condition is applied on the wedge surface, and an extrapolation boundary condition is used at the right outflow boundary.

## 3. Results and Discussion

Numerical simulations are performed to investigate the oblique detonation structures under three activation energy conditions: low activation energy (C=1.0), high activation energy (C=1.1) and very low activation energy (C=0.9). A comparative analysis is conducted to examine the influence of activation energy variation on the wave-front morphology and detonation instability. Table 1 summarizes the theoretical predictions of the incoming flow and thermodynamic properties of the oblique detonation, which serve as reference values for comparison with the numerical results.

**Table 1. The key flow parameters**

| Parameters | Values |
|---|---|
| Freestream velocity (m/s) | 2473.4 |
| Freestream temperature (K) | 851.5 |
| Freestream pressure (kPa) | 42.5 |
| Wedge angle (deg) | 25 |
| Normal velocity of freestream or detonation velocity (m/s) | 2074.3 |
| Tangential velocity (m/s) | 1331.1 |
| Oblique detonation angle (deg) | 57.5 |
| Oblique shock wave angle (deg) | 39.9 |
| Sound velocity of detonation products (m/s) | 1130.6 |
| Sound velocity behind oblique shock wave (m/s) | 914.5 |
| Static temperature behind oblique shock wave (K) | 1550 |

### 3.1 Low Activation-Energy Case (C=1.0)

Figures 1 presents the pressure and vector-flux contours of the oblique detonation for the low activation energy case. The horizontal and vertical axes denote the grid indices. As shown in the pressure contours, the oblique detonation structure can be clearly divided into three regions: an induction zone, an overdriven detonation zone, and a transverse wave region. A transverse wave emerges downstream of overdriven region and propagates in the downstream direction. Figure 2 shows magnified views of the pressure and vector-flux contours in the induction and overdriven regions. The location of the deflagration wave can be clearly identified. The vector-flux field further indicates that, at low activation energy, the leading shock and the heat release remain well coupled, and the rarefaction waves are very weak (no significant red contours are observed).

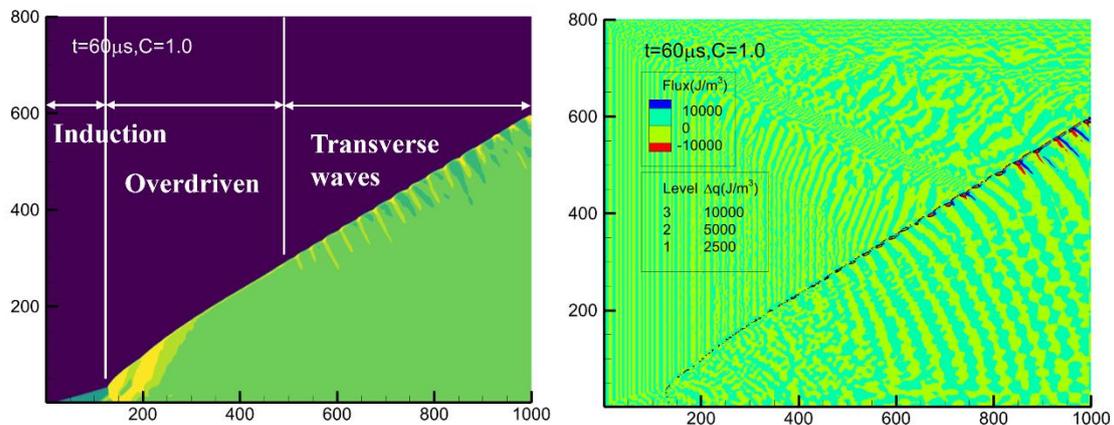

Figure 1. Pressure and vector-flux contours

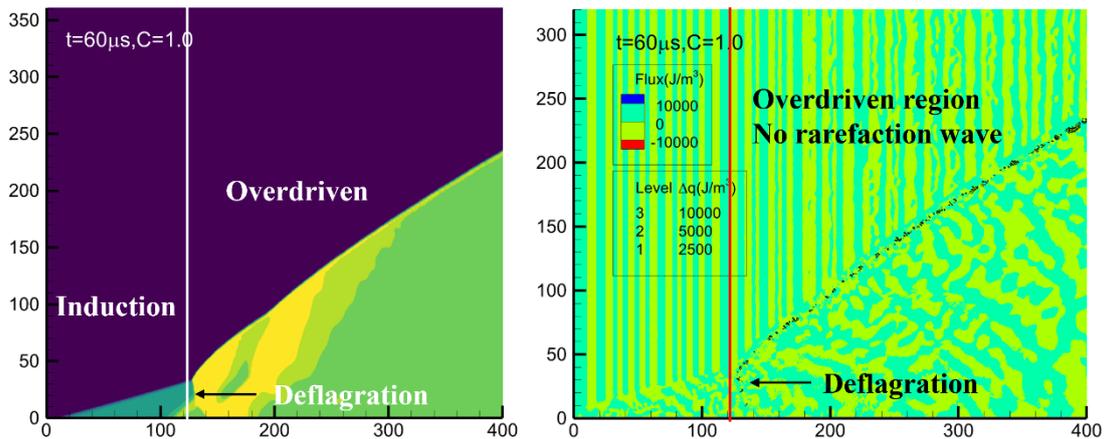

Figure 2. Pressure and vector-flux contours in the induction and overdriven region

Figure 3 shows enlarged pressure and vector-flux contours in the transverse wave region. As seen from the pressure field, a transverse wave originates near point A and propagates downstream, gradually strengthening. Figure 4 presents a further magnified view of the vector-flux distribution at point A. It is evident that partial decoupling occurs along the oblique detonation front: a portion of the heat release remains coupled with the leading shock (visible as the overlap of blue contours with the black line), while another portion couples with an unsteady rarefaction wave (overlap of red contours with black line). The unsteady rarefaction wave reduces the total energy per unit volume, such that although chemical heat release occurs, it does not produce a pressure rise. This marks the onset of transverse wave formation at the oblique detonation front. Upstream of point A lies the overdriven detonation region, where the shock and reaction zone remain tightly coupled and no transverse waves develop.

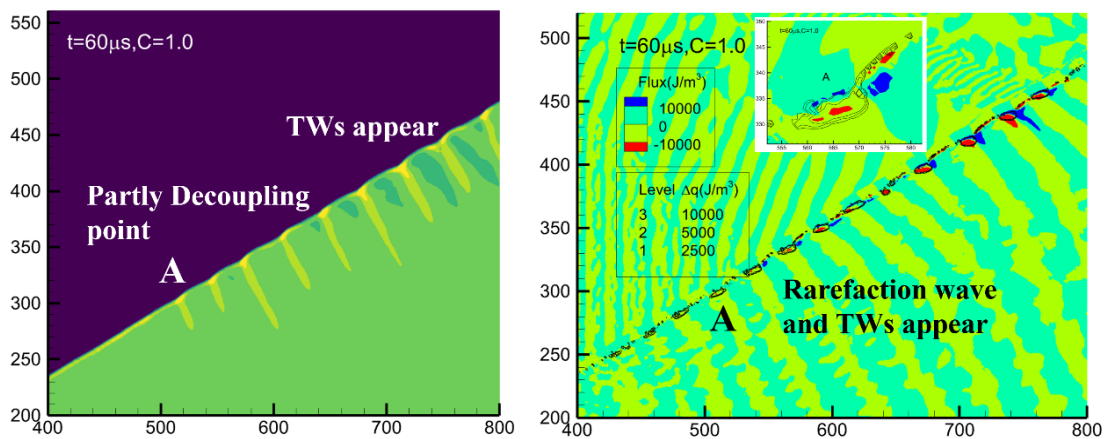

Figure 3. Pressure and vector-flux contours in the transverse wave region

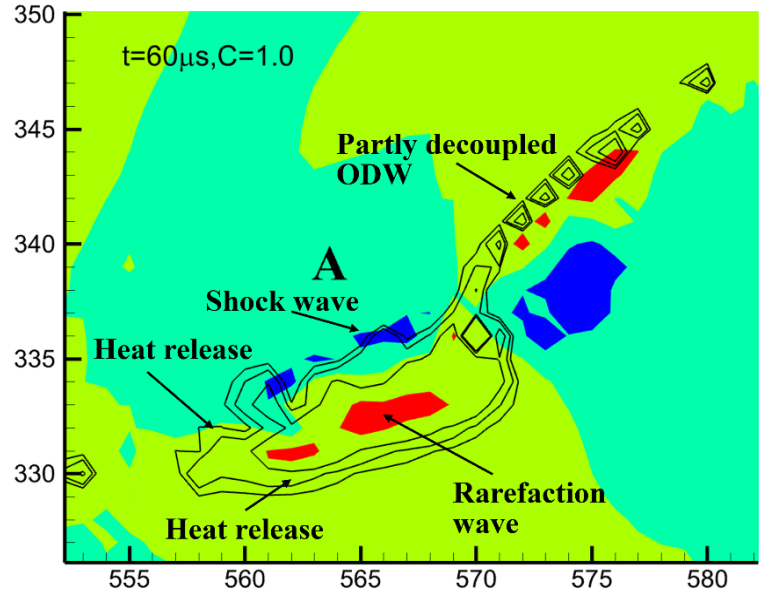

Figure 4. Enlarged vector-flux contour at point A

Figures 5 and 6 present the pressure and vector flux contours in the transverse wave region, together with the detailed structure of the transverse wave. From the pressure field alone, one may infer the existence of both upward-running and downward-running transverse waves. However, the vector flux contours clearly show that only a downward-propagating transverse wave is present. This transverse wave induces a strong unsteady rarefaction (highlighted in red), indicating significant local energy extraction from the flow.

Figure 7 shows the positions of the transverse wave at two time instants separated by 0.1μs interval. The downstream motion of the transverse wave is evident. The measured tangential propagation speed of the transverse wave is approximately 2380m/s. According to Table 1, the theoretical tangential flow speed behind the oblique detonation front is 1331m/s, implying a transverse wave propagation speed of about 1049m/s, relative to the flow, which is very close to the sound speed of the detonation products. Using the sound speed immediately behind the oblique shock of 914.5m/s, the corresponding transverse wave Mach number is approximately Mach 1.15.

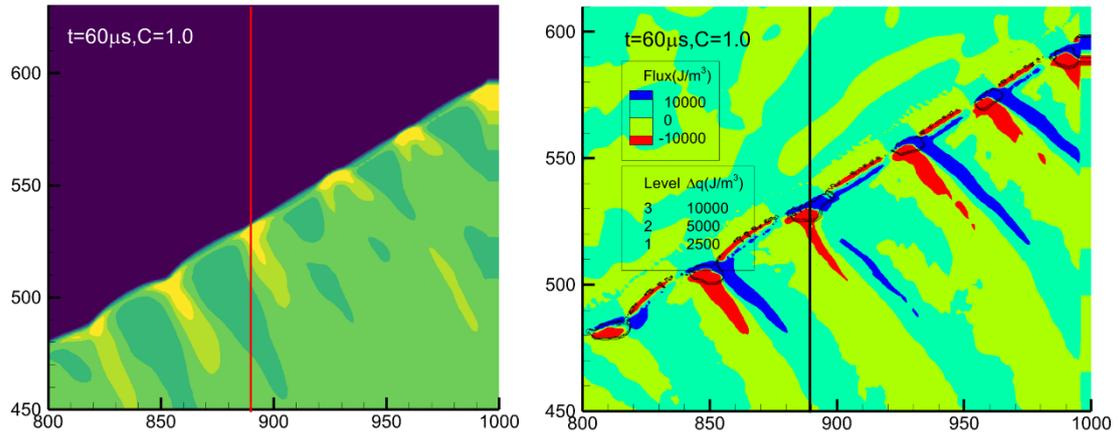

Figure 5. Enlarged vector-flux contour in the transverse wave region

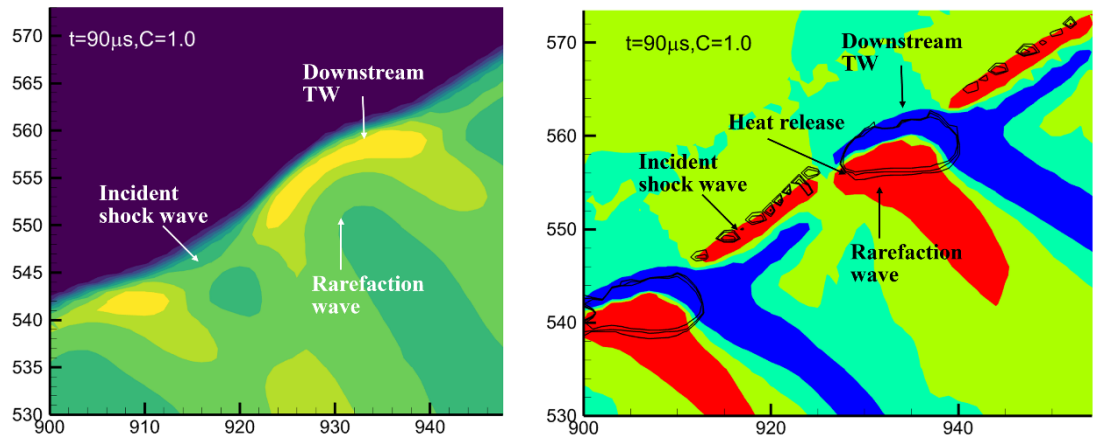

Figure 6. Detailed structure of the transverse wave

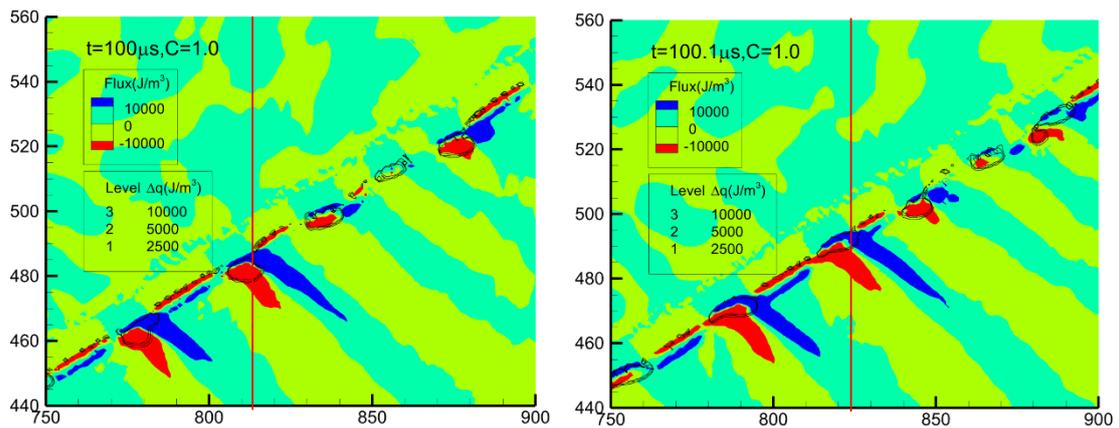

Figure 7. Propagation speed of the transverse wave

## 3.2 High Activation-Energy Case (C=1.1)

Figures 8 presents the pressure and vector flux contours of the oblique detonation under the high activation energy condition. In contrast to the low activation energy case, the transverse wave region is directly connected to the overdriven detonation zone. The transverse waves propagate upward and eventually merges into the overdriven detonation front. Figure 9 shows the detailed wave structure of the traverse wave, and Figure 9 provides vector flux snapshots separated by 1μs. It is observed that the upward-propagating transverse wave has an absolute tangential velocity of approximately 300m/s. Considering the local tangential flow velocity, the transverse wave propagation speed relative to the flow is approximately 1431m/s, corresponding to a Mach number of about Mach 1.26.

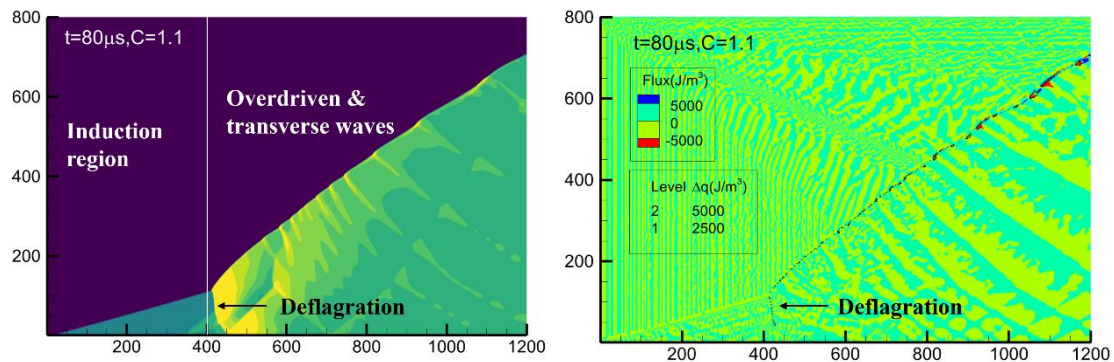

Figure 8. Pressure and vector-flux contours

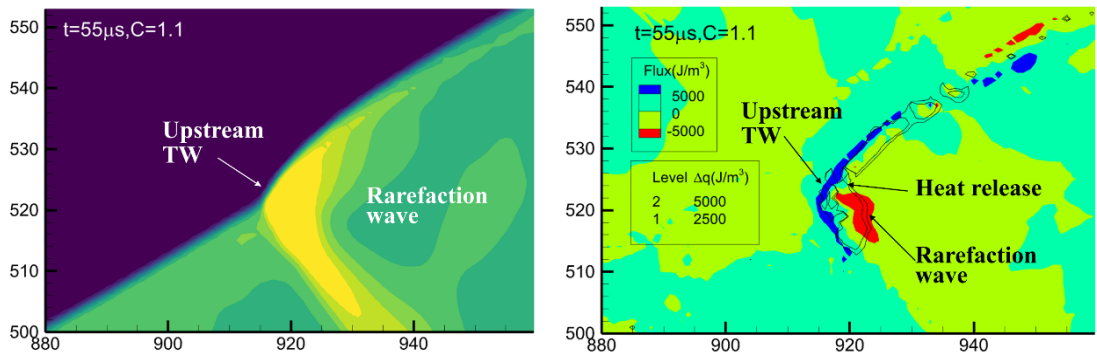

Figure 9. Detailed structure of the transverse wave

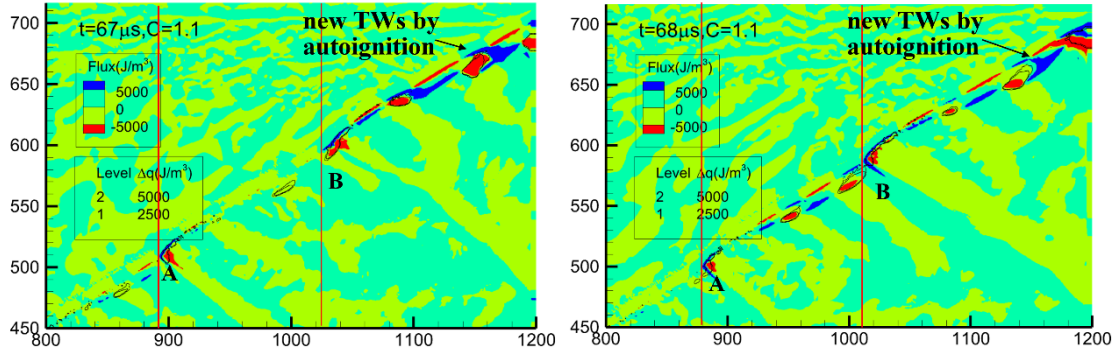

Figure 10. Propagation speed of the transverse wave

Under the high activation energy condition, the transverse wave propagates upward. As a result, the downstream transverse wave gradually moves away from the outlet boundary. To sustain a stable oblique detonation, the nonlinear reactive Euler system periodically generates new transverse waves in the downstream region through an auto-ignition mechanism. The period of new transverse wave formation corresponds to the ignition delay time of the detonable mixture. Table 2 lists the ignition delay times at different activation energy levels and temperatures. The theoretical post-oblique-shock static temperature is approximately 1550K. For C=1.1, the ignition delay time at 1550K is 6.2 μs. Figure 10 shows the pressure contours at four different times. A periodic generation of new transverse waves is clearly observed, driven by auto-ignition mechanism. The average period is approximately 7 μs，which agrees well with the predicted ignition delay time of 6.2 μs at 1550 K.

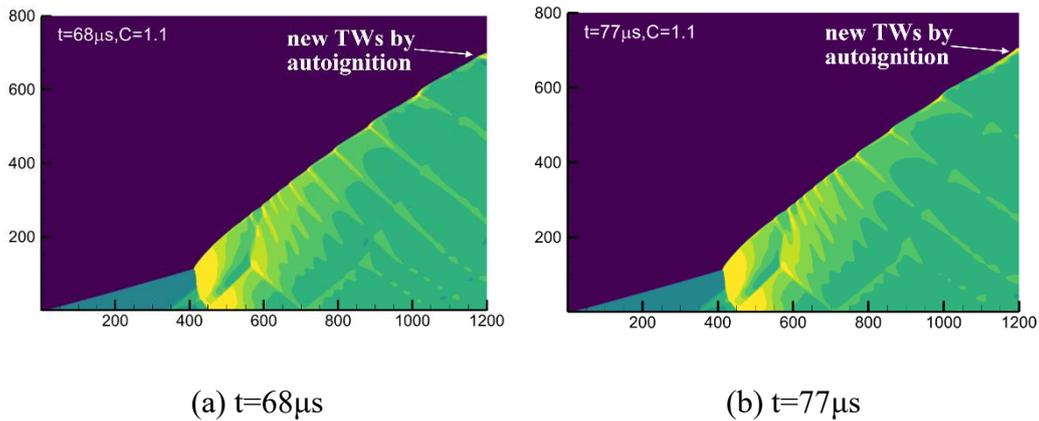

(a) t=68μs                                    (b) t=77μs

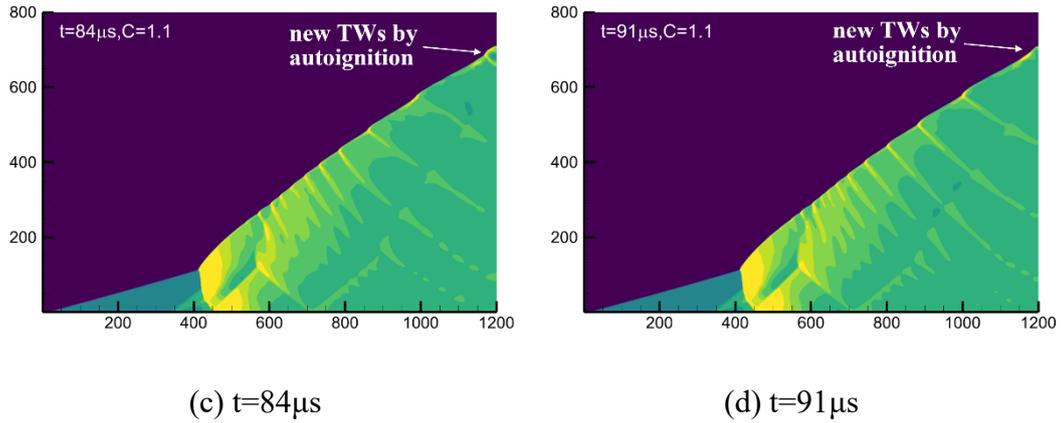

(c) t=84μs                  (d) t=91μs

Figure 11. Formation of new transverse waves by auto-ignition

**Table 2. Ignition delay time under different activation energy**

| Activation energy | Temperature (K) | Ignition delay time (μs) |
|:---:|:---:|:---:|
| C=0.9 | 1550 | 0.6 |
|  | 1600 | 0.4 |
| C=1.0 | 1550 | 1.9 |
|  | 1600 | 1.4 |
| C=1.1 | 1550 | 6.2 |
|  | 1600 | 4.3 |

### 3.3 Very Low Activation-Energy Case (C=0.9)

Figure 12 presents the pressure and vector-flux contours of the oblique detonation under the very low activation energy condition. It is evident that no transverse wave forms along the oblique detonation front. This behavior results from the extremely rapid heat release, which produces only very weak rarefaction effects. Consequently, the heat release remains tightly coupled with the leading shock, preventing transverse wave generation and maintaining a fully stable oblique detonation front.

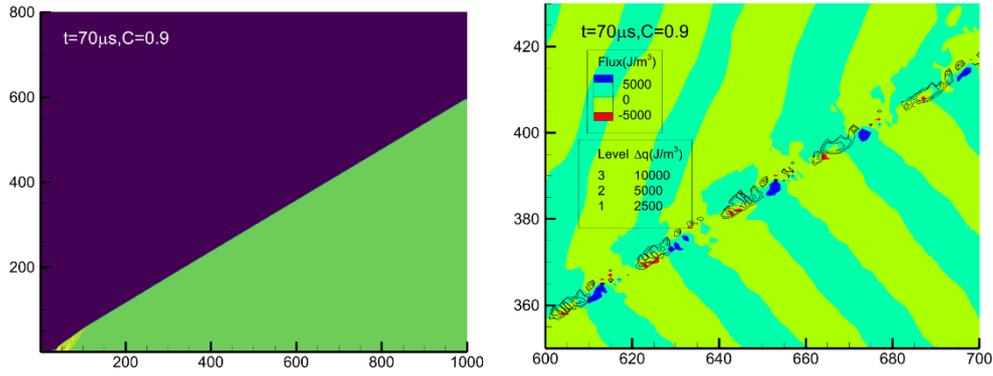

Figure 12. Pressure and vector-flux contours under very low activation energy

## 3.4 Unsteady Initiation Process of Oblique Detonation (C=1.1)

In this subsection, the unsteady initiation process of the oblique detonation is examined using the high activation energy case as an example. As shown in Table 2, the ignition delay time of the stoichiometric hydrogen-air mixture at high activation energy is 6.2 μs; therefore, it can be anticipated that no significant chemical heat release occurs within the first 5 μs. Figure 13 illustrates the transient initiation of the oblique detonation wave. At t=5 μs, no chemical reaction is observed. At t=6μs, auto-ignition begins to occur. By t=8μs, the heat release from auto-ignition generates a shock inside the oblique shock wave layer, which then propagates outward. At t=10μs, the shock induced by combustion couples with the oblique shock, marking the establishment of the oblique detonation. This behavior is consistent with theoretical predictions.

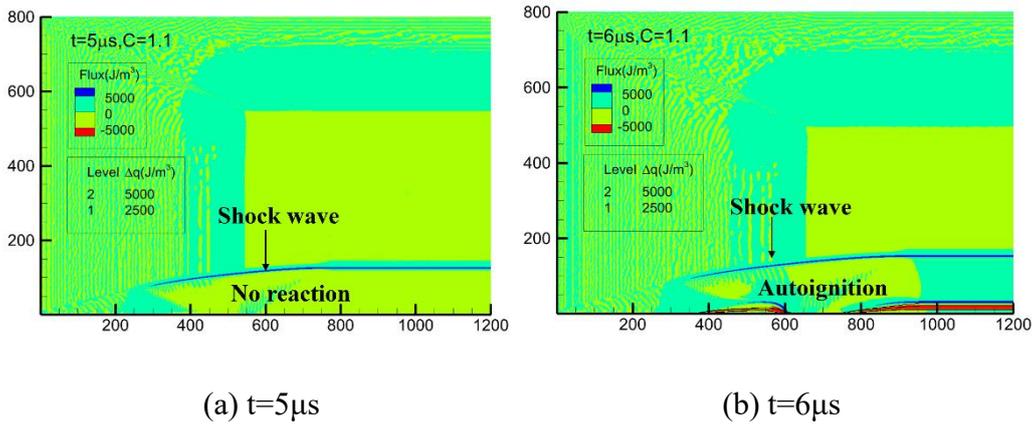

(a) t=5μs             (b) t=6μs

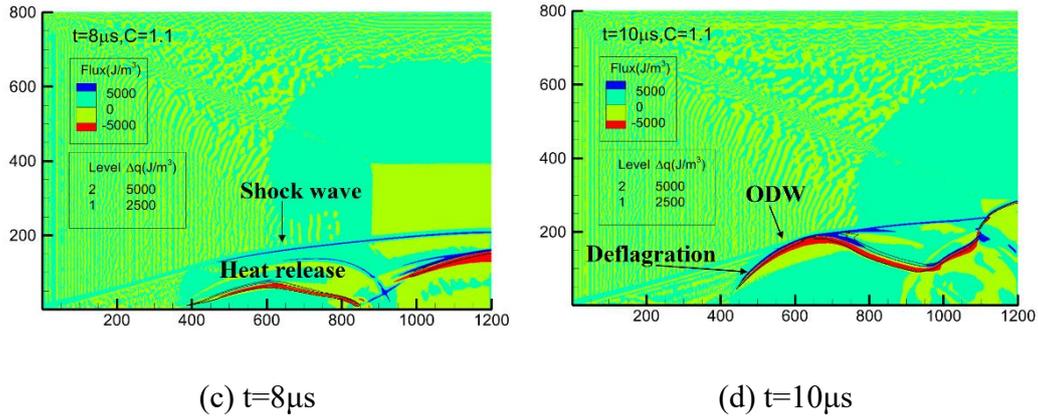

(c) t=8μs                                    (d) t=10μs

Figure 13. Unsteady initiation process of the oblique detonation

Figure 14 presents the reaction progress variable $Z$, temperature field, vector-flux contours, and a schematic of the steady oblique detonation structure. As illustrated in Figure 14(d), the propagation mechanism of the oblique detonation can be summarized as follows. First, within the induction region, auto-ignition occurs in the lowest layer of the mixture adjacent to the wedge surface. The auto-ignition heat-release front is nearly parallel to the oblique shock. The resulting pressure rise propagates upstream, forming a deflagration wave that is almost perpendicular to the incoming flow within the induction zone. This deflagration wave consumes the mixture in the upper portion of the induction region. The pressure increase generated by the deflagration wave strengthens the oblique shock, producing an overdriven detonation wave, which subsequently decays toward the Chapman-Jouguet (CJ) detonation state. These results indicate that auto-ignition of the mixture layer closest to the wedge wall is the key physical mechanism responsible for establishing the stable oblique detonation.

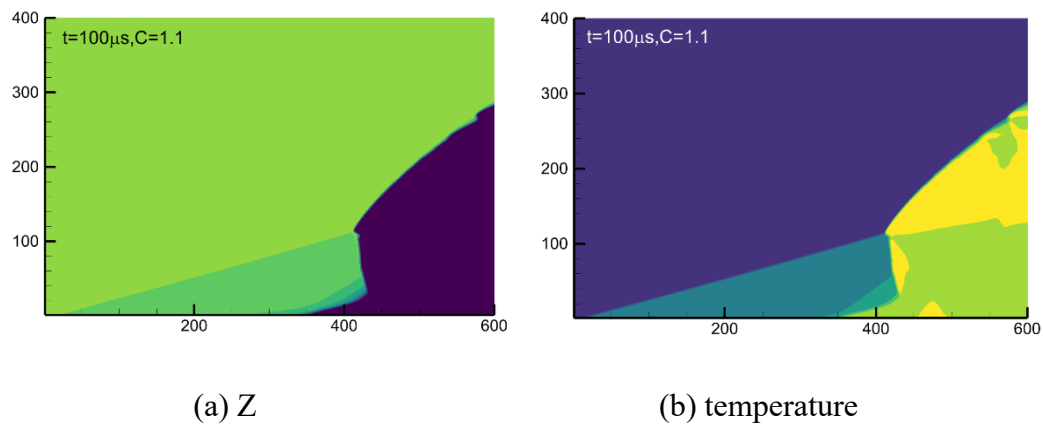

(a) Z                                        (b) temperature

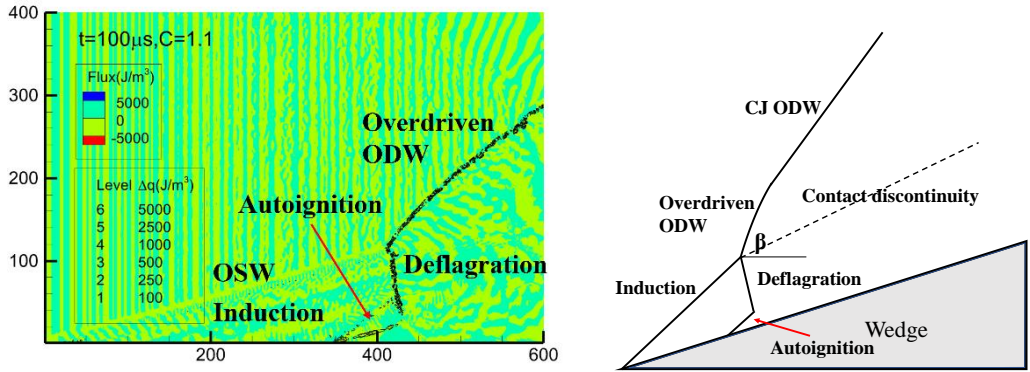

|                |                |
| :------------: | :------------: |
| (c) vector-flux | (d) structure |

Figure 14. Structure of the stable oblique detonation wave

## 4. Conclusion

In this study, two-dimensional numerical simulations of stoichiometric hydrogen-air oblique detonation waves were performed using the conservative Euler equations and a one-step global reaction model. The activation energy was varied to investigate the transverse wave propagation mechanisms. The wedge angle was 25°, and the freestream conditions were $T$=851.5K, u=2473.4m/s, and p=42.5kPa. The results show that the oblique detonation wave consists of three distinct regions: an induction zone, an overdriven detonation zone, and a transverse wave region. Under low activation energy, a downward-propagating transverse wave develops. This wave originates from detonation front decoupling due to the gradual attenuation of the detonation strength, and its propagation speed is approximately equal to the sum of the tangential flow velocity and the sound speed of the detonation products. In contrast, under high activation energy, an upward-propagating transverse wave forms and travels toward the upstream direction, merging into the overdriven detonation zone. The relative propagation speed of this wave is approximately 300m/s. To sustain a stable oblique detonation, new transverse waves are periodically generated by auto-ignition downstream of the last transverse wave. The generation period matches the mixture ignition delay time. For downward-propagating transverse waves, new waves are generated upstream; while for upward-propagating transverse waves, new waves are generated downstream. Their mechanism is different. Thus, upward-propagating and downward-propagating transverse waves cannot coexist simultaneously.